\documentclass{amsart}
\begin{document} 
\begin{flushright}
solv-int/9904003
\end{flushright}
\title[B\"{a}cklund 
transformations]{B\"{a}cklund 
transformations for 
many-body systems related to KdV}
\author{A.N.W.~Hone}
  \address{Dipartimento di Fisica `E. Amaldi',
Universita degli Studi `Roma Tre', Roma, Italia}
\author{V.B.~Kuznetsov}
  \address{Department of Applied Mathematics,  
          University of Leeds, 
          Leeds LS2 9JT, UK}
\author{O.~Ragnisco}
  \address{Dipartimento di Fisica `E. Amaldi',
Universita degli Studi `Roma Tre', Roma, Italia}
\renewcommand{\theequation}
{\arabic{section}.\arabic{equation}} 

\begin{abstract} 
We present B\"{a}cklund transformations 
(BTs) with parameter 
for certain classical integrable $n$-body 
systems, namely the many-body generalised  
H\'{e}non-Heiles, Garnier and Neumann  
systems. Our construction makes use of 
the fact that all these systems may be  
obtained as particular  
reductions (stationary or restricted flows) 
of the KdV hierarchy; alternatively they 
may be considered as examples of the reduced $sl(2)$ 
Gaudin magnet. The BTs 
provide exact time-discretizations of the 
original (continuous) systems, preserving 
the Lax  matrix and hence all integrals of motion, 
and satisfy the {\it spectrality} property 
with respect to the B\"{a}cklund parameter.
\end{abstract} 
\maketitle 

\section{Introduction} 
\noindent
B\"{a}cklund transformations (BTs) 
are an important aspect of the 
theory of integrable 
systems which have traditionally 
been studied in the context of 
evolution equations. However, more recently 
there has been much interest in discrete 
systems or integrable mappings 
\cite{sympl, veselov}. Within the 
modern approach to separation of 
variables (reviewed by Sklyanin in \cite{sklyan})
this has led to the study of  
BTs for finite-dimensional 
Hamiltonian systems \cite{kuskly}. 
The latter are canonical transformations 
including a B\"{a}cklund parameter $\lambda$, 
and apart from being interesting integrable 
mappings in their own right they also 
lead to separation of variables when $n$ such 
mappings are applied to an 
integrable system with $n$ degrees of freedom.  
The sequence of B\"{a}cklund 
parameters $\lambda_{j}$ together with a set 
of conjugate variables $\mu_{j}$ 
constitute the separation variables, and 
satisfy a new property called 
{\it spectrality} introduced in \cite{kuskly}. 

We  proceed to develop these ideas 
with some new examples of BTs for $n$-body 
systems, namely the many-body 
generalisation of the case (ii) 
integrable H\'{e}non-Heiles system,  the 
Garnier system and the Neumann system 
on the sphere (see \cite{eekt}).
It is known that the case (ii) 
H\'{e}non-Heiles system is equivalent to 
the stationary flow of fifth-order KdV 
\cite{fordy}, while the Garnier and 
Neumann systems may be obtained as 
restricted flows of the KdV hierarchy \cite{zeng}.   
Thus we derive  BTs for these systems 
by reduction of 
the standard BT for KdV, which arises 
from the Darboux-Crum transformation 
\cite{crum} for Schr\"{o}dinger operators. 
The restriction of the Darboux transformation 
to the stationary flows of the modified (mKdV) 
hierarchy has been discussed in \cite{poiss}. 

In the following section we describe how 
the reduction works in general,  
before specialising these 
considerations to each particular system 
and presenting the associated 
generating function for the BT. 
We note that these systems are examples 
of the reduced Gaudin magnet \cite{eekt}, so that we have
the following Lax matrix 
\begin{equation} 
L(u)=\sum_{j=1}^{n}\frac{\ell_{j}}{u-a_{j}}+B(u), 
\qquad \ell_{j}=\left( 
\begin{array}{cc} S_{j}^{3} & 
S_{j}^{-} \\ S_{j}^{+} & -S_{j}^{3} \end{array} 
\right)  \label{eq:gaulax} 
\end{equation} 
where (up to scaling) 
the $S_{j}$ satisfy $n$  independent 
copies of the standard $sl(2)$ algebra: 
\begin{equation} 
\{ S_{j}^{3},S_{k}^{\pm} \} 
=\pm2\delta_{jk}S_{k}^{\pm}, 
\qquad  \{ S_{j}^{+},S_{k}^{-} \} 
=4\delta_{jk}S_{k}^{3}.   \label{eq:alg} 
\end{equation}  
For the H\'{e}non-Heiles 
and Garnier systems the matrix $B(u)$ is 
respectively quadratic and linear in the 
spectral parameter $u$, while for the 
Neumann system it is independent of 
$u$ and turns out to be constant due to 
the constraint that the particles lie on the 
sphere (hence the Poisson algebra 
(\ref{eq:alg}) must be modified by Dirac 
reduction). 

We have constructed the BT for the (non-reduced)
$sl(2)$ Gaudin magnet with quasi-periodic boundary
condition in \cite{gaud}, 
while some preliminary results on the 
classical Garnier system 
and two-body H\'{e}non-Heiles system
first appeared in \cite{us}. 

\section{Classical integrable systems and KdV} 
\setcounter{equation}{0} 
\subsection{Restricting the BT}
\noindent
As is well known, the Darboux-Crum transformation 
\cite{crum} consists of mapping the 
Schr\"{o}dinger operator 
$\partial_{t}^{2}+V-\lambda$ to another 
operator $\partial_{t}^{2}+\tilde{V}-\lambda$ 
by factorizing the former and then 
reversing the order of factorisation. 
Given an eigenfunction $\phi$ satisfying 
$$ 
(\partial_{t}^{2}+V-\lambda)\phi=0 
$$ 
we may set $y=(\log[\phi])_{t}$ and then 
\begin{equation} 
V=-y_{t}-y^{2}+\lambda, \qquad 
\tilde{V}=y_{t}-y^{2} +\lambda; \label{eq:miura} 
\end{equation} 
for $\lambda=0$ this is just the Miura map for 
KdV. Also given another 
eigenfunction $\psi$ of the Schr\"{o}dinger 
operator with potential $V$ 
for a different spectral parameter $u$ we have 
$$ 
(\partial_{t}^{2}+V-u)\psi=0, \qquad 
(\partial_{t}^{2}+\tilde{V}-u)\tilde{\psi}=0 
$$ 
where the transformation to 
the new eigenfunction $\tilde{\psi}$
and its derivative may be given in matrix form as 
\begin{equation} 
\left(\begin{array}{c} \tilde{\psi}_{t} \\ 
\tilde{\psi} \end{array} \right) = k
\left(\begin{array}{cc} -y & y^{2} + u -  \lambda \\ 
 1 & -y \end{array} \right) 
\left(\begin{array}{c} \psi_{t} \\ 
\psi  \end{array} \right) 
\label{eq:gauge} 
\end{equation} 
for any constant $k$.  From (\ref{eq:miura}) 
follows the standard formula for 
the Darboux-B\"{a}cklund 
transformation of KdV, 
$\tilde{V}=V+2(\log[\phi])_{tt}$. 

For what follows it will also be necessary to 
consider a product of eigenfunctions 
for a Schr\"{o}dinger operator with potential 
$V$ and eigenvalue $u$, 
$$ 
f=\psi\psi'
$$ 
with Wronskian 
$\psi_{t}\psi '-\psi\psi_{t}'=2m$. 
It is well known 
that $f$ satisfies the Ermakov-Pinney 
equation \cite{pinney} 
\begin{equation} 
ff_{tt}-\frac{1}{2}f_{t}^{2}+2(V-u)f^{2}+2m^{2}=0. 
\label{eq:pinney} 
\end{equation}  
If we now transform $\psi$ and $\psi'$ according 
to (\ref{eq:gauge}) then we find a new 
product of eigenfunctions 
$\tilde{f}=\tilde{\psi} \tilde{\psi}'$  
satisfying the 
same Ermakov-Pinney equation but with 
$V$ replaced by $\tilde{V}$, given explicitly by  
\begin{equation} 
\tilde{f}=(\lambda-u)^{-1}\frac{(Z^{2}-m^{2})}{f}, 
\quad Z=\frac{1}{2}f_{t}-yf,   
\label{eq:trans} 
\end{equation} 
where we have set $k^{2}=(\lambda-u)^{-1}$ to 
ensure that the transformed eigenfunctions have 
the same Wronskian $2m$. It is also 
straightforward to show that,  in terms of 
$\tilde{f}$, the quantity $Z$ can be written as
$Z=-\frac{1}{2}\tilde{f}_{t}-y\tilde{f}$ 
(see \cite{exact}).
 
We can now  describe how 
this transformation restricts 
to the finite-dimensional Hamiltonian systems 
presented below. The systems are expressed 
in variables $(q_{j},p_{j})$
which appear in the Lax matrix (\ref{eq:gaulax}) 
via the identification \cite{ku,eekt}
$$ 
S_{j}^{3}=p_{j}q_{j}, \quad 
S_{j}^{-}=-p_{j}^{2}+\frac{m_{j}^{2}}{q_{j}^{2}}, 
\quad 
S_{j}^{+}=q_{j}^{2}. 
$$ 
For H\'{e}non-Heiles and Garnier the 
non-vanishing Poisson 
brackets are the standard ones 
$\{p_{j},q_{k}\}=\delta_{jk}$ 
which provide a realization   of the 
algebra (\ref{eq:alg}); for the Neumann 
system on the sphere the bracket must 
be modified by Dirac reduction. 

All of the systems 
are Liouville integrable, and thus have 
a complete set of Hamiltonians 
in involution, but for these purposes we 
concentrate on the Hamiltonian $h$ generating 
the flow corresponding to $t$ 
above (in KdV theory this is usually  
denoted $x$, the spatial variable). 
For this flow the Lax equation $L_{t}=[N,L]$ 
is the compatibility condition for the linear 
system 
\begin{equation} 
L(u)\Psi=v\Psi, \quad \Psi_{t}=N\Psi; 
\quad  
N=\left(\begin{array}{cc} 0 & u-V(q_{j},p_{j}) \\  
1 & 0 \end{array} \right). \label{eq:linsys} 
\end{equation} 
Observe that the second part of the linear system 
is just a Schr\"{o}dinger equation for the 
potential $V$; for Neumann and Garnier this is 
a function of $(q_{j},p_{j})$ for 
$j=1,\ldots,n$, while for H\'{e}non-Heiles 
there is an extra pair of conjugate variables  
$(q_{n+1},p_{n+1})$ such that $V\equiv q_{n+1}$. 

The equations of motion generated by this 
Hamiltonian take the form  $q_{j,t}=p_{j}$ 
and 
\begin{equation}  
p_{j,t}=q_{j,tt}=(a_{j}-V(q_{k},p_{k}))q_{j}-
\frac{m_{j}^{2}}{q_{j}^{3}} \label{eq:pini} 
\end{equation} 
for $j=1,\ldots,n$; for H\'{e}non-Heiles 
there are also equations for 
$q_{n+1}$ and $p_{n+1}=q_{n+1,t}$. 
The important thing to observe is that 
(\ref{eq:pini}) is equivalent to the fact that 
$S_{j}^{+}=q_{j}^{2}$ satisfies the 
Ermakov-Pinney  equation (\ref{eq:pinney}) 
corresponding to a Schr\"{o}\-dinger equation with 
potential $V$ and eigenvalue $a_{j}$. Thus 
to obtain a B\"{a}cklund transformation for 
these many-body systems we simply apply 
a Darboux-Crum transformation 
(\ref{eq:miura}) to the potential 
$V=V(q_{j},p_{j})$ to obtain 
$\tilde{V}=V(\tilde{q}_{j},\tilde{p}_{j})$, 
and then we know that the solutions 
of the Ermakov-Pinney equation must 
transform according to (\ref{eq:trans}).  
By this procedure we may 
explicitly construct the BT 
for  the many-body systems
below (or for any restricted flow of KdV), 
and it is then simple to calculate the 
generating function $F(q_{j},\tilde{q}_{j})$ 
of this canonical 
transformation, such that 
$$ 
dF=\sum_{j}
(p_{j}dq_{j}-\tilde{p}_{j}d\tilde{q}_{j}). 
$$ 

The discrete Lax equation for the BT,  
$$ 
ML=\tilde{L}M
$$   
where 
$\tilde{L}=L(\tilde{q_{j}}, \tilde{p_{j}};u)$, 
is necessary to ensure  the 
preservation of the 
spectral curve $\det(v-L(u))=0$  
(so that 
all the Hamiltonians in involution 
 are  preserved). 
This follows 
immediately from the 
properties of the Darboux-Crum 
transformation, since we know that 
the vector $\Psi$ in the linear system 
(\ref{eq:linsys}) must transform 
according to (\ref{eq:gauge}), and hence 
we may take (setting $k=1$) 
\begin{equation} 
M=\left( 
\begin{array}{cc} -y & y^{2} + u -  \lambda \\ 
 1 & -y \end{array} \right). 
\label{eq:emm} 
\end{equation} 
Of course we must 
determine $y$ as a function of 
the dynamical variables. 
In the Garnier and H\'{e}non-Heiles 
cases it turns out that the potential 
depends on coordinates only, $V=V(q_{j})$, 
and so by adding the two equations in 
(\ref{eq:miura}) we obtain  
$$ 
y(q_{j}, \tilde{q_{j}})=\pm
\sqrt{\lambda-\frac{1}{2}(V+\tilde{V})}; 
$$ 
to obtain the correct continuum limit 
of the discrete dynamics it is necessary to 
take the negative branch of the square root 
(see \cite{us, exact}). For the Neumann system 
$V$ depends on both coordinates and 
momenta, so the above does not yield 
$y(q_{j}, \tilde{q_{j}})$. 

There is another way of writing $L$ 
which arises more naturally via 
reduction from the zero curvature 
representation of the KdV 
hierarchy \cite{eekt,fordy, zeng}, viz 
$$ 
L(u)=\left(\begin{array}{cr}
\frac{1}{2}\Pi_{t}     
&  -\frac{1}{2}\Pi_{tt} +(u-V)\Pi\\ 
\Pi  & -\frac{1}{2}\Pi_{t} 
\end{array} \right)
$$ 
where 
\begin{equation} 
\Pi(u)=\sum_{j=1}^{n}\frac{q_{j}^{2}}{u-a_{j}} 
+\Delta(u). 
\label{eq:other} 
\end{equation}
$\Delta$ is a polynomial in $u$ fixing the dynamical term $B$ 
in (\ref{eq:gaulax}); we shall present the appropriate $\Delta$ 
and $B$ in each case below.  
Clearly the $t$ derivatives of 
$\Pi$ can be rewritten using the equations 
of motion to yield (\ref{eq:gaulax}).  

Finally if we write the 
(hyper-elliptic) spectral curve as 
$$ 
v^{2}=R(u)
$$
then it is easy to check that the spectrality property \cite{kuskly}
is satisfied for these systems, in the sense that defining the 
conjugate variable to $\lambda$ by 
$$ 
\mu=-2\frac{\partial F}{\partial \lambda} 
$$ 
we find that 
$$ 
L(\lambda)\Omega=\mu\Omega 
$$ 
with eigenvector $\Omega=(y,1)^{T}$, 
or in other words $\mu^{2}=R(\lambda)$ so that 
$(\lambda,\mu)$ is a point on the 
spectral curve. Note  that 
(as for the examples in \cite{gaud, kuskly}) 
this eigenvector spans the kernel of $M$, 
$$
M(\lambda)\Omega=0. 
$$
We can also write $y$ explicitly 
in terms of both the old and the 
new variables related by the BT, thus: 
\begin{equation} 
y(q_{j},p_{j})=
\frac{\Pi_{t}(\lambda)+2\mu}{2\Pi(\lambda)}, \qquad  
y(\tilde{q_{j}},\tilde{p_{j}}) =
-\frac{(\tilde{\Pi}_{t}(\lambda)-2\mu)}
{2 \tilde{\Pi}(\lambda)}; 
\label{eq:ynewold} 
\end{equation} 
clearly we denote $\tilde{\Pi}(\lambda)=
\Pi(\tilde{q_{j}},\tilde{p_{j}};\lambda)$. 
 
\subsection{Generalised H\'{e}non-Heiles system} 
\noindent
For the many-body generalisation  of 
case (ii) integrable H\'{e}non-Heiles 
system, the  Hamiltonian
generating the $t$ flow takes the form 
$$ 
h=\frac{1}{2}\sum_{j=1}^{n+1}p_{j}^{2} 
+q_{n+1}^{3}+q_{n+1} 
\left(\frac{1}{2}\sum_{j=1}^{n} 
q_{j}^{2}+c\right) 
-\frac{1}{2}\sum_{j=1}^{n} 
\left(a_{j}q_{j}^{2}+ 
\frac{m_{j}^{2}}{q_{j}^{2}}\right). 
$$ 
The original case (ii) integrable 
H\'{e}non-Heiles 
system corresponds to $n=1$ with 
$c=m_{j}=a_{j}=0$. 
The link between stationary 
fifth-order KdV and the type (ii) 
system 
was noted by Fordy in \cite{fordy}, although 
this was anticipated 
in work of Weiss \cite{weiss}, 
who used Painlev\'{e} analysis to derive a BT 
and associated linear problem 
(a similar result also appears in \cite{new}).  
None of these authors wrote a BT 
explicitly as a canonical 
transformation with parameter, 
although (without parameter) 
this was done for a 
non-autonomous version in 
\cite{thesis}.

For the Lax matrix $L$ of the 
generalised $(n+1)$-body H\'{e}non-Heiles 
system 
we have $\Delta=-16u-8q_{n+1}$ 
so that the extra term $B(u)$ is given by 
$$ 
B=\left(\begin{array}{cc} 
-4p_{n+1} & E \\ 
-16u-8q_{n+1} & 4p_{n+1} 
\end{array}\right), 
\quad  
E=-16u^{2}+8q_{n+1}u 
-4q_{n+1}^{2}-\sum_{j=1}^{n} 
q_{j}^{2}-4c. 
$$ 
The equations of motion for $h$ imply that the 
squares of the first $n$ coordinates 
$q_{j}^{2}$ satisfy the Ermakov-Pinney 
equation (\ref{eq:pinney}) for $m=m_{j}$ 
with 
$$ 
V=q_{n+1} 
$$  
and eigenvalue $a_{j}$. Thus the BT for the 
system can be calculated directly by 
applying the Darboux-Crum 
transformation to $V=q_{n+1}$, to yield 
$\tilde{V}=\tilde{q}_{n+1}$, and 
applying (\ref{eq:trans}) to each $q_{j}^{2}$ 
for $j=1,\ldots,n$. 

After some calculation the generating function for 
this canonical transformation is found to be 
$$ 
F(q_{j},\tilde{q}_{j};\lambda)=\sum_{j=1}^{n} 
\left( 
Z_{j}+\frac{m_{j}}{2}\log\left[ 
\frac{Z_{j}-m_{j}}{Z_{j}+m_{j}}\right]\right) 
+ 
\frac{16}{5}y^{5}+4(q_{n+1}+\tilde{q}_{n+1})y^{3} 
$$  
$$ 
+\left(2q_{n+1}^{2}+2q_{n+1}\tilde{q}_{n+1} 
+2\tilde{q}_{n+1}^{2}+\frac{1}{2}\sum_{j=1}^{n} 
(q_{j}^{2}+\tilde{q}_{j}^{2})+2c\right)y, 
$$ 
where we have found it convenient  
to use the quantities $Z_{j}(q_{j},\tilde{q}_{j})$ and 
$y(q_{j},\tilde{q}_{j})$ defined by 
\begin{equation} 
Z_{j}^{2}=m_{j}^{2}+ 
(\lambda-a_{j})q_{j}^{2}\tilde{q}_{j}^{2}, 
\label{eq:zeb} 
\end{equation} 
and 
$$ 
y=-\sqrt{\lambda-\frac{1}{2}(q_{n+1}+\tilde{q}_{n+1})}. 
$$ 

In order to check the spectrality property,  
we have explicitly found  that 
the eigenvalue of $L(\lambda)$ with 
eigenvector $\Omega=(y,1)^{T}$ can be written as 
$$ 
\mu(q_{j},\tilde{q}_{j};\lambda)  
=-\sum_{j=1}^{n}\frac{Z_{j}}{\lambda-a_{j}}- 
\frac{1}{y}\frac{\partial F}{\partial y},  
$$ 
which precisely equals 
$-2\frac{\partial F}{\partial  \lambda}$ as 
required. 

\subsection{Garnier system}  
\noindent
For the Garnier system the $t$ flow is 
generated by the Hamiltonian 
$$ 
h=\frac{1}{2}\sum_{j=1}^{n}p_{j}^{2}
+\frac{1}{2}(\sum_{j=1}^{n}q_{j}^{2})^{2}
-\frac{1}{2}\sum_{j=1}^{n}\left(a_{j}q_{j}^{2}+ 
\frac{m_{j}^{2}}{q_{j}^{2}}\right). 
$$ 
This differs from the traditional Garnier system 
as in \cite{us, wojc, zeng} 
by the inclusion of  extra inverse square terms. 
The Newton equations for the $q_{j}$ are  
$$ 
q_{j,tt}+2(\sum_{k}q_{k}^{2})q_{j}= 
a_{j}q_{j}-\frac{m_{j}^{2}}{q_{j}^{3}},  
$$ 
so clearly for the standard restricted flows of KdV 
\cite{zeng}, when $m_{j}=0$, each 
$q_{j}$ is an eigenfunction 
of a Schr\"{o}dinger operator 
with potential 
$$
V=2\sum_{j}q_{j}^{2} 
$$ 
and eigenvalue $a_{j}$, while in general 
$q_{j}^{2}$ is a product of eigenfunctions 
satisfying the Ermakov-Pinney equation 
\cite{pinney} for $m=m_{j}$.  

The Lax matrix of the Garnier system has 
$\Delta=1$, so $L$ takes the form 
(\ref{eq:gaulax}) with
$$ 
B=\left(\begin{array}{cc}  
0 & u-\sum_{j}q_{j}^{2} \\ 
1 & 0 \end{array}\right). 
$$  
Applying the Darboux-Crum 
transformation we obtain a new 
potential 
$$ 
\tilde{V}=2\sum_{j}\tilde{q}_{j}^{2},  
$$ 
and the corresponding BT induced on the 
Garnier system is equivalent to gauging 
$L$ by the matrix $M$ of the form 
(\ref{eq:emm}) with 
$$ 
y=-\sqrt{\lambda-\sum_{j} 
(q_{j}^{2}+\tilde{q}_{j}^{2})}.
$$ 

Finally we can calculate the generating function 
for this BT, which may be written as follows:
$$ 
F(q_{j},\tilde{q}_{j};\lambda)=\sum_{j=1}^{n} \left( 
Z_{j}+\frac{m_{j}}{2}\log\left[ 
\frac{Z_{j}-m_{j}}{Z_{j}+m_{j}}\right]\right) 
-\frac{1}{3}y^{3},
$$ 
where $y(q_{j},\tilde{q}_{j})$ is as above 
and $Z_{j}$ is given by the same 
expression (\ref{eq:zeb}) as for 
H\'{e}non-Heiles. In \cite{us} we derived 
this generating function for the special 
case $m_{j}=0$ when the logarithm terms 
do not appear. To check spectrality 
we notice that $L(\lambda)$ has eigenvalue 
$$ 
\mu(q_{j},\tilde{q}_{j};\lambda)  
=-\sum_{j=1}^{n}\frac{Z_{j}}{\lambda-a_{j}} +y 
$$ 
with eigenvector $\Omega$, and so we see that
$\mu=-2\frac{\partial F}{\partial  \lambda}$. 

\subsection{Neumann system on the sphere} 
\noindent
For the Neumann system the $t$ flow is 
generated by  
$$ 
h=\frac{1}{2}\sum_{j=1}^{n}p_{j}^{2} 
-\frac{1}{2}\sum_{j=1}^{n}\left(a_{j}q_{j}^{2}+ 
\frac{m_{j}^{2}}{q_{j}^{2}}\right). 
$$ 
Once again this has extra inverse square terms 
compared with the standard Neumann system 
\cite{ragn, wojc}. The Poisson bracket for 
this system is modified by constraining 
the particles to lie on a sphere, so that 
\begin{equation} 
(q,q)\equiv\sum_{j}q_{j}^{2}=const, \qquad 
(q,p)\equiv\sum_{j}q_{j}p_{j}=0 
\label{eq:constr}  
\end{equation} 
which results in the non-vanishing 
Dirac brackets 
\begin{equation} 
\{p_{j},q_{k}\}=\delta_{jk}- 
\frac{q_{j}q_{k}}{(q,q)}, \quad 
\{p_{j},p_{k}\}=\frac{q_{j}p_{k}-q_{k}p_{j}}{(q,q)}. 
\label{eq:dirac} 
\end{equation} 
With this bracket the Hamilton equations are 
$q_{j,t}=p_{j}$ and (\ref{eq:pini}) with 
$$ 
V=(q,q)^{-1} \sum_{j}\left( p_{j}^{2}+a_{j}q_{j}^{2} 
-\frac{m_{j}^{2}}{q_{j}^{2}}\right). 
$$ 

The Lax matrix for the Neumann system 
arises by setting $\Delta=0$, which 
in (\ref{eq:gaulax}) gives the following matrix $B$: 
$$ 
B=\left(\begin{array}{cc} 
0 & (q,q) \\ 
0 & 0 \end{array}\right). 
$$ 
In fact if we start from the linear system 
(\ref{eq:linsys}) and leave $V$ unspecified 
then  (\ref{eq:pini}) as well as the constraint  
$(q,q)_{t}=0$ are consequences of the Lax 
equation, and together these are sufficient to 
determine the form of $V$; this is also how 
the equations for the 
constrained Neumann system arise in 
a Lagrangian approach \cite{ragn}. 

Given that the phase space is now 
degenerate with two Casimirs given by 
(\ref{eq:constr}), 
it would appear that the standard sort 
of generating function will no longer be 
appropriate for describing a BT. It turns out 
that we can apply  the Darboux-Crum 
transformation  
as before, and transform the quantities $q_{j}^{2}$ 
according to (\ref{eq:trans}). In this way we obtain 
new variables $\tilde{q}_{j}(q_{k},p_{k})$ and 
$\tilde{p}_{j}(q_{k},p_{k})$, which are naturally 
written with the use of the quantity $y(q_{k},p_{k})$ 
given by the first formula in (\ref{eq:ynewold}); 
on the Lax matrix this transformation arises 
by gauging with $M$ as in (\ref{eq:emm}).   
Similarly the transformation can be inverted to give 
$q_{j}(\tilde{q}_{k},\tilde{p}_{k})$ and 
$p_{j}(\tilde{q}_{k},\tilde{p}_{k})$ written in 
terms of $y(\tilde{q}_{k},\tilde{p}_{k})$ given 
by the right hand formula of (\ref{eq:ynewold}). 

However, it would still be nice to write a 
generating function for this transformation. 
We have found that if we formally take 
$$ 
F(q_{j},\tilde{q}_{j};\lambda)=\sum_{j=1}^{n} \left( 
Z_{j}+\frac{m_{j}}{2}\log\left[ 
\frac{Z_{j}-m_{j}}{Z_{j}+m_{j}}\right] 
+\frac{1}{2}y(q_{j}^{2}-\tilde{q}_{j}^{2})\right)
$$ 
with $Z_{j}$ given by (\ref{eq:zeb}) as usual, 
and regard $y$ as a sort of Lagrange 
multiplier (independent of the coordinates 
and $\lambda$), then we do indeed 
obtain the correct expressions 
$$ 
p_{j}=\frac{\partial F}{\partial q_{j}}, \qquad 
\tilde{p}_{j}=-\frac{\partial F}{\partial\tilde{q}_{j}}, 
$$ 
but these contain $y$ which is unspecified. If we 
then require that the constraints (\ref{eq:constr}) 
are preserved under the BT applied from old to new 
variables or vice-versa, then in either direction 
the constraints are preserved if and only if $y$ 
satisfies a quadratic equation with solution 
given respectively by the formulae 
(\ref{eq:ynewold}). Alternatively if we 
require spectrality then second 
component of the equation 
$L(\lambda)\Omega=\mu\Omega$  gives 
$$ 
\mu(q_{j},\tilde{q}_{j};\lambda)  
=-\sum_{j=1}^{n}\frac{Z_{j}}{\lambda-a_{j}}= 
-2\frac{\partial F}{\partial  \lambda} 
$$ 
as required, 
while the first component gives 
(after making use of the formula 
(\ref{eq:zeb}) and the BT) 
$$ 
\mu=-\sum_{j=1}^{n}\frac{Z_{j}}{\lambda-a_{j}} 
+\frac{1}{y}\sum_{j}(q_{j}^{2}-\tilde{q}_{j}^{2}). 
$$ 
Hence spectrality requires that the second term 
vanishes, and so the first constraint 
(\ref{eq:constr}) is preserved; the preservation 
of the second constraint is then an  
algebraic consequence of the BT.  

Thus we see that for this BT we can write 
the new variables as functions of the 
old and vice-versa, but a formula for 
$y(q_{j},\tilde{q}_{j};\lambda)$ is lacking. 
Also this discretization of the Neumann 
system is apparently new, since it is exact 
(preserving the Lax matrix for the continuous
system) unlike the Veselov or Ragnisco 
discretizations discussed in \cite{ragn}. 

\section{Conclusions} 
\noindent
It would also be interesting to 
look at BTs with parameter in the 
non-autonomous case \cite{thesis}, where deformation 
with respect to the B\"{a}cklund parameter 
would probably have to be 
introduced (corresponding to the 
associated isomonodromy problem). 

\section{Acknowledgements} 
\noindent
ANWH thanks the Leverhulme Trust for  providing  
a Study Abroad Studentship in Rome, 
and is grateful to J.~Harnad and Y.~Suris 
for useful conversations.   
VBK acknowledges the support from the EPSRC and
the support from Istituto Nazionale di Fisica Nucleare
for his visit to Rome.
The authors would also like to thank the 
organisers of the meeting {\it Integrable 
Systems: Solutions and Transformations} 
in Guardamar, Spain (June 1998) where 
some of this work was carried out.

\end{document}